\documentstyle[12pt]{article}

\makeatletter
\def\eqnarray{\stepcounter{equation}\let\@currentlabel=\theequation
\global\@eqnswtrue
\global\@eqcnt\z@\tabskip\@centering\let\\=\@eqncr
$$\halign to \displaywidth\bgroup\@eqnsel\hskip\@centering
  $\displaystyle\tabskip\z@{##}$&\global\@eqcnt\@ne
  \hfil$\displaystyle{\hbox{}##\hbox{}}$\hfil
  &\global\@eqcnt\tw@ $\displaystyle\tabskip\z@
  {##}$\hfil\tabskip\@centering&\llap{##}\tabskip\z@\cr}
\def\@sect#1#2#3#4#5#6[#7]#8{\ifnum #2>\c@secnumdepth
    \def\@svsec{}\else
    \refstepcounter{#1}\edef\@svsec{\csname the#1\endcsname.\hskip 1em
    }\fi
    \@tempskipa #5\relax
    \ifdim \@tempskipa>\z@
    \begingroup #6\relax
    \@hangfrom{\hskip #3\relax\@svsec}{\interlinepenalty \@M #8\par}
    \endgroup
    \csname #1mark\endcsname{#7}\addcontentsline
    {toc}{#1}{\ifnum #2>\c@secnumdepth \else
     \protect\numberline{\csname the#1\endcsname}\fi
           #7}\else
    \def\@svsechd{#6\hskip #3\@svsec #8\csname #1mark\endcsname
          {#7}\addcontentsline
          {toc}{#1}{\ifnum #2>\c@secnumdepth \else
     \protect\numberline{\csname the#1\endcsname}\fi
           #7}}\fi
     \@xsect{#5}}
\def\label#1{\@bsphack\if@filesw {\let\thepage\relax
   \xdef\@gtempa{\write\@auxout{\string
   \newlabel{#1}{{\thesection.\@currentlabel}{\thepage}}}}}\@gtempa
   \if@nobreak \ifvmode\nobreak\fi\fi\fi\@esphack}
\def\@eqnnum{(\thesection.\theequation)}
\def\section{\setcounter{equation}{0} \@startsection {section}{1}{\z@}
{-3.5ex
   plus -1ex minus -.2ex}{2.3ex plus .2ex}{\Large\bf}}
\newcount\@minsofar
\newcount\@min
\newcount\@cite@temp
\def\@citex[#1]#2{%
\if@filesw \immediate \write \@auxout {\string \citation {#2}}\fi
\@tempcntb\m@ne \let\@h@ld\relax \def\@citea{}%
\@min\m@ne%
\@cite{%
  \@for \@citeb:=#2\do {\@ifundefined {b@\@citeb}%
    {\@h@ld\@citea\@tempcntb\m@ne{\bf ?}%
    \@warning {Citation `\@citeb ' on page \thepage \space
    undefined}}%
{\@minsofar\z@ \@for \@scan@cites:=#2\do {%
  \@ifundefined{b@\@scan@cites}%
    {\@cite@temp\m@ne}
    {\@cite@temp\number\csname b@\@scan@cites \endcsname \relax}%
\ifnum\@cite@temp > \@min% select the next one to list
    \ifnum\@minsofar = \z@
      \@minsofar\number\@cite@temp
      \edef\@scan@copy{\@scan@cites}\else
    \ifnum\@cite@temp < \@minsofar
      \@minsofar\number\@cite@temp
      \edef\@scan@copy{\@scan@cites}\fi\fi\fi}\@tempcnta\@min
  \ifnum\@minsofar > \z@ % some more
    \advance\@tempcnta\@ne
    \@min\@minsofar
    \ifnum\@tempcnta=\@minsofar %   Number follows previous--hold on
    to it
      \ifx\@h@ld\relax
        \edef \@h@ld{\@citea\csname b@\@scan@copy\endcsname}%
    \else \edef\@h@ld{\ifmmode{-}\else--\fi\csname
    b@\@scan@copy\endcsname}%
      \fi
    \else \@h@ld\@citea\csname b@\@scan@copy\endcsname
          \let\@h@ld\relax
  \fi % no more
\fi}%
\def\@citea{,\penalty\@highpenalty\,}}\@h@ld}{#1}}
\def\appendixname{Appendix}
\def\appendix{\par
  \def\pre@section{\appendixname{}}
  \setcounter{section}{1}
  \@addtoreset{equation}{section}
  \def\thesection{\Alph{section}}
  \def\theequation{\arabic{equation}}}

\def\appendix{\par
  \def\pre@section{\appendixname{}}
  \setcounter{section}{1}
  \@addtoreset{equation}{section}
  \def\thesection{\Alph{section}}
  \def\theequation{\arabic{equation}}}

\makeatother

\def\s{\sigma}

\def\l{\lambda}
\def\e{\epsilon}

\def\ds{\displaystyle}
\def\be{\begin{equation}}
\def\ee{\end{equation}}
\def\beq{\begin{eqnarray}}
\def\eeq{\end{eqnarray}}

\begin{document}
\begin{center}
\bf {Quantum Spin Chains and Riemann Zeta Function with Odd Arguments}
\end{center}
\phantom{a}

\vspace{1.5cm}

\centerline{H.E. Boos\footnote{
E-mail: boos@mx.ihep.su}}
\centerline{
Institute for High Energy Physics}
\centerline{ Protvino, 142284, Russia}

\phantom{a}

\vspace{0.5cm}

\phantom{a}

\centerline{V.E. Korepin \footnote{E-mail:
korepin@insti.physics.sunysb.edu}}

\centerline{C.N.~Yang Institute for Theoretical Physics}
\centerline{State University of New York at Stony Brook}
\centerline{Stony Brook, NY 11794--3840, USA}

\vspace{1.5cm}

\vskip2em
\begin{abstract}
\noindent
Riemann zeta function is an important object of number theory.
It was also  used for description of disordered systems in statistical
mechanics. We show that Riemann zeta function is also useful for the
description of
integrable model. We study XXX  Heisenberg spin 1/2 anti-ferromagnet.
We evaluate a  probability of formation of a
ferromagnetic string in the anti-ferromagnetic ground state in
thermodynamics limit.
We prove that  for short strings the probability can be expressed
in terms of Riemann zeta function with odd arguments.

\end{abstract}

\newpage

\section{Introduction}

 Riemann zeta function for $Re (s) > 1 $ can be defined as follows:

\be
\zeta (s) = \sum_{n=1}^{\infty}\frac{1}{n^s}
\label{zeta}
\ee

It also can be represented as a product with respect to all prime
numbers $p$
\be
\zeta (s) = \prod_{p}  ( 1-p^{-s}  ) ^{-1}
\ee

It can be analytically continued in the whole complex plane of $s$.
It has only one pole, at $s=1$ and it has 'trivial' zeros at $s=-2n$ (
$n>1$ is an integer).
The famous Riemann hypothesis  \cite{R}  states that nontrivial zeros
belong to the straight line
$Re (s) = 1/2$.
Riemann zeta function is useful for study of distribution of prime
numbers on the
real axis  \cite{TIT}. The values of Riemann zeta function at special
points were studied in
\cite{za} , \cite{bbbl} .
 At even values of its argument  zeta function can be expressed in
 terms of
powers of $\pi$.
The values of Riemann zeta function at odd arguments
provide infinitely many different irrational numbers \cite{tr} .
% Mathematicians believe
%that at odd values of its argument $\zeta (n)$ is an irrational
%number [ this was proven for
%$\zeta (3)$].
Riemann zeta function  plays an important role, not only in pure
mathematics
%\cite{za} ,\cite{bbbl}
but also  theoretical physics. Some Feynman diagrams in
quantum field theory can be expressed in terms of $\zeta (n)$,
see, for example, \cite{KR} .
It appears also in string theory \cite{string}.
In statistical mechanics Riemann zeta function  was used for the
description of chaotic systems.
This is large field with many publications.
Important contributions to this field were made by  Berry, Connes,
Julia, Kac, Keating,
Knauf, Odlyzko,  Pitkanen, Polya,  Ruelle, Sarnak and Zagier.
% This isdefinitely incomplete list.
%We apologize to the authors, whose names we did not mention.
 One can find more information and citation  on  the following web
 cite
{http://www.maths.ex.ac.uk/~mwatkins/} .

%\cite{con} , \cite{kna}

We argue that  $\zeta (n)$ is also important for exactly solvable
models.
One of the most famous integrable models is the Heisenberg  XXX spin chain.
This model was first suggested by Heisenberg  \cite{Heis} in 1928
and
solved by Bethe  \cite{B} in 1931 .
Since that time it found multiple applications in solid state physics
and
statistical mechanics.
Recently the XXX spin chain was used for study of the entanglement in
quantum computations  \cite{comp}.

The Hamiltonian of the XXX spin chain can be written like this

\be
H = \sum_{i=1}^N \>
(\s^x_i\s^x_{i+1}\; + \;\s^y_i\s^y_{i+1}\; +\; \s^z_i\s^z_{i+1}\;-1\;)
\label{H}
\ee
Here $N$ is the length of the lattice and  $\s^x_i,\s^y_i,\s^z_i$ are
Pauli  matrices. We consider thermodynamics limit , when $N$ goes to infinity.
 The sign in front of the Hamiltonian indicates that we
are considering the
anti-ferromagnetic case. We consider periodic boundary conditions.
Notice that this Hamiltonian  annihilates the ferromagnetic state [
all spins up].

The construction of the anti-ferromagnetic ground state wave
function $|AFM>$  can be credited to
Hulth\'{e}n  \cite{H}. An important
correlation function was defined in \cite{KIEU}.
It was called the emptiness formation probability

$$
P(n) = <AFM|\prod_{j=1}^n P_j|AFM>
$$
%$P_j=\frac{(\s^z_j+1)}{2}$
where $P_j= (1+\s^z_j)/2$ is a projector on the state with spin up in
$j$th lattice site.
Averaging is over the
anti-ferromagnetic ground state. It describes the
probability of formation of a
ferromagnetic string of the length $n$ in the
anti-ferromagnetic background  $|AFM>$ .
%Our main results is $P(4)$.
In this paper we shall first study short strings ( $n$ is small),
in the end we shall discuss
long distance asymptotics ( at finite temperature).
The four first values of the emptiness-formation probability look as
follows:
\beq
&{\ds P(1)\;=\;{1\over 2}}\;=\;0.5,&\label{P1}\\
&{\ds P(2)\;=\;{1\over 3} (1\; -\; \ln{2})}\;=\;0.102284273,&
\label{P2}\\
&{\ds P(3)\;=\;{1\over 4}\; - \;\ln{2}\;+\;{3\over 8}\;\zeta(3)}\;=
\;0.007624158,&
\label{P3}\\
&{\ds P(4)\;=\;{1\over 5}\; -\; 2\ln{2} \; + \; {173\over 60}
\,\zeta(3)
\; -\; {11\over 6} \,\zeta(3)\, \ln{2}\;-\; {51\over 80} \, \zeta^2(3)
\;
}&
\nonumber\\
&{\ds
 -\; {55\over 24}\,\zeta(5)
\; + \; {85\over 24}\,\zeta(5)\,  \ln{2}\;=\;0.000206270}&
%\nonumber\\
%&\quad&
\label{P4}
\eeq
 Let us comment.
The value of  $P(1)$ is evident from the symmetry, $P(2)$
can be extracted from the explicit expression of the ground
state energy \cite{H}.
$P(3)$  can be extracted from the results of M.Takahashi \cite{T1}
on the calculation of the nearest neighbor correlation. 
It was confirmed in paper \cite{DI}.
One should also mention independent calculation of $P(3)$ in
\cite{BGSS}.
One can  express $P(3)$ in terms of next to the nearest neighbor
correlation
%Namely,
%as was shown in \cite{BGSS} one can  express $P(3)$ in terms of
%next to the nearest neighbor correlation via the formula

\be
<\;S^z_{i}S^z_{i+2}\;>\;=\;2\,P(3)\;-\;2\,P(2)\;+\;{1\over 2}\,P(1)
\label{G2a}
\ee

The calculation of $P(3)$ and
$P(4)$ is discussed in this paper.

{\bf The expression above for $P(4)$ is our main result here.}

The plan of the paper is as follows. In the next section we discuss
some main steps of the calculation of $P(3)$ and $P(4)$.
The thermodynamics of $P(n)$ for the non-zero temperature
is briefly discussed in section 3.
Then we summarize the results in the conclusion.

\section{General discussion of the calculation of $P(3)$ and $P(4)$}

There are several different approaches to investigate $P(n)$:

\begin{itemize}

\item
{\bf representation of correlation functions as determinants of
Fredholm integral operators}
 described in detail in the book
\cite{KIB}

\item
{\bf the vertex operator approach} developed by the RIMS group
\cite{JMMN}

\end{itemize}

One can also mention the application of connection with other
correlation functions,
for instance, the correlation function $<AFM|S^z_{i}S^z_{i+n}|AFM>$.

We shall use the integral representation obtained by
Korepin, Izergin, Essler and Uglov \cite{KIEU} in framework
of the vertex operator approach at the zero magnetic field:
\be
P(n)=\int_C {d\lambda_1\over 2\pi i\lambda_1}
\int_C {d\lambda_2\over 2\pi i\lambda_2}\ldots
\int_C {d\lambda_n\over 2\pi i\lambda_n}
\prod_{a=1}^n (1+{i\over\lambda_a})^{n-a}
({\pi\lambda_a\over\sinh{\pi\lambda_a}})^n
\prod_{1\le j<k\le n}{\sinh{\pi(\lambda_k-\lambda_j)}
\over\pi(\lambda_k-\lambda_j-i)}.
\label{intPn}
\ee
The contour $C$ in each integral goes parallel to the real axis
with the imaginary part between \\
$0$ and $-i$.

Recently such formula was  generalized by de~Gier and Korepin in paper
\cite{GK}
to the case, where averaging is done over arbitrary Bethe state [ with
no strings ]
instead of anti-ferromagnetic state.

Let us describe in general the strategy we used in order to come
to the answers (\ref{P3}) and (\ref{P4}).
The integral formula (\ref{intPn}) can be easily represented
as follows:
\be
P(n)=
\prod_{j=1}^n\int_{C}{d\lambda_j\;\over 2\pi i }\;
U(\l_1,\ldots,\l_n)\;T(\l_1,\ldots,\l_n)
\label{intPna}
\ee
where
\be
U(\l_1,\ldots,\l_n)\;=\;\pi^{{n(n+1)\over 2}}\>
{\prod_{1\leq k < j\leq n}\sinh{\pi(\lambda_j-\lambda_k)}
\over
\prod_{j=1}^n\sinh^n{\pi\lambda_j}
}
\label{U_n}
\ee
and
\be
T(\l_1,\ldots,\l_n)\;=\;
{\prod_{j=1}^n\l_j^{j-1}(\lambda_j+i)^{n-j}\over
\prod_{1\leq k < j\leq n}(\lambda_j-\lambda_k-i)
}
\label{T_n}
\ee

As appeared we can make a lot of simplifications without taking
integrals but using some simple observations.
First of all, let us note that the function $U(\l_1,\ldots,\l_n)$
is antisymmetric in respect to transposition of any pair of
integration variables, say, $\l_j$ and $\l_k$. This simple
observation turns out to be very useful because
\be
\prod_{j=1}^n\int_{C}{d\lambda_j\;\over 2\pi i }\;
U(\l_1,\ldots,\l_n)\;S(\l_1,\ldots,\l_n)\;=\;0
\label{int}
\ee
if the function $S$ is symmetric for at least one pair of $\l$-s.

The next observation is also trivial, namely, we can try to reduce
the power of denominator in (\ref{T_n}) using simple algebraic
relations like
\be
{1\over x(x+a)}\;=\;{1\over a x}\;-\;{1\over a (x+a)}.
\label{xa}
\ee

Combining these two simple observations one can reduce integration
functions for $P(3)$ to a sum of terms with denominators of power 2
and for $P(4)$ to a more complicated sum of terms with denominators
of power not higher than 3.

In order to calculate the integrals one can close the contours in the
complex plane by the infinite semi-circles either in upper half-plane
or in the lower half-plane not changing the integrals. Then it is
possible to apply Cauchy theorem using the following formulae
\beq
&{\ds
\oint_{C_l} {dz\over 2\pi i}\; {f(z)\over \sinh^3{\pi z}}\;=\;-{1\over
2\pi}\;
{(1-{1\over\pi^2}
{\partial^2
\over{{\partial\e}^2}})}_{\e\rightarrow 0}\;f(i\,l+\e)}&
\label{c3l}\\
&{\ds
\oint_{C_l}  {dz\over 2\pi i}\; {f(z)\over \sinh^4{\pi z}}\;=
\;-{2\over 3\pi^2}\;
{({\partial\over{{\partial\e}}}
-{1\over 4\pi^2}{\partial^3\over{{\partial\e}^3}})}_{\e\rightarrow 0}
\;
f(i\,l+\e)
}&\label{c4l}
\eeq
for the cases $n=3$ and $n=4$ respectively where $C_l$ is a small
contour surrounding the point $i\, l$ with an integer $l$ in
anti-clockwise direction.

Then the integrals can be expressed in terms of the differential
operator acting on some functions. For instance, for the case
$n=3$
\be
\int_{C}{d\lambda_1\;\over 2\pi i }
\int_{C}{d\lambda_2\;\over 2\pi i }
\int_{C}{d\lambda_3\;\over 2\pi i }U(\l_1,\l_2,\l_3)
F(\l_1,\l_2,\l_3)\;=\;D\;\tilde F(\e_1,\e_2,\e_3)
\label{IS}
\ee
where $D$ is the differential operator
$$
%\be
D\;=\;-{\pi^3\over 8}\;{
(1-{1\over\pi^2}{\partial^2\over{{\partial\e_1}^2}})
(1-{1\over\pi^2}{\partial^2\over{{\partial\e_2}^2}})
(1-{1\over\pi^2}{\partial^2\over{{\partial\e_3}^2}})
}_{\e_1,\e_2,\e_3\rightarrow 0}
$$
\be
\sinh{\pi (\e_2-\e_1)}\,\sinh{\pi (\e_3-\e_1)}\,\sinh{\pi (\e_3-\e_2)}
\label{D}
\ee
and
\be
\tilde F(\e_1,\e_2,\e_3)\;=\;\sum_{l_1=0}^{\infty}(-1)^{l_1}
\sum_{l_2=0}^{\infty}(-1)^{l_2}\sum_{l_3=0}^{\infty}(-1)^{l_3}
F(i\,l_1\,+\,\e_1,i\,l_2\,+\,\e_2,i\,l_3\,+\,\e_3)
\label{F}
\ee
Here all three contours were closed in the upper half-plane but
in real calculations it turns out to be more convenient to close
some of them in another direction taking into consideration appearance
of an additional sign.

It is not difficult to get generalization of these formulae to the
case $n=4$. So the problem is reduced to the calculation of sums
like (\ref{F}), expanding the result into the series in powers
of $\e$-s and applying the differential operator $D$. This procedure
is straightforward but can be rather tedious especially for the case
$n=4$. Proceeding in this way we can come to the results (\ref{P3})
and
(\ref{P4}).

Let us note that both of these final answers appeared
to be expressed in terms of the logarithmic function and the Riemann
zeta function of odd arguments and do not depend on polylogarithms
in spite of the fact that  polylogarithm $\mbox{Li}_4(1/2)$
appeared in the intermediate stage of calculation. All coefficients
before
those functions in (\ref{P1}-\ref{P4}) are rational.
Also they do not contain any powers of $\pi$ which could be considered
as Riemann zeta functions of even arguments.

{\bf Our conjecture is that the final answer
for any $P(n)$ will also be expressed in terms of logarithm
$\ln{2}$ and Riemann zeta functions $\zeta(k)$ with odd integers $k$
and with rational coefficients.}

\section{Thermodynamics of $P(n)$}

If we had the exact answer for $P(n)$ for any $n$ we could calculate
an asymptotics of $P(n)$ when $n$ tends to infinity.
Unfortunately, for a moment we can not do this because we have
$P(n)$ only for $n=1,2,3,4$. Nevertheless we can discuss a possible
behavior of $P(n)$ with $n\rightarrow\infty$ using some other
arguments.

For non-zero temperature one can conclude that the asymptotics of
the partition function in thermodynamic limit is as follows
\be
Z\;=\;<\;e^{{H\over kT}}\;>\;\sim\;e^{{Nf\over kT}}
\label{Z}
\ee
where $f$ is the free energy per site and $N$ is the length of the
chain, it was evaluated in \cite{TS}, \cite{TSK} and \cite {T}.
 In fact, for $P(n)$ the  $n$ neighboring  spins are frozen. Therefore
one has the asymptotics of $P(n)$ when $n$ tends to infinity

\be
P(n)\;=\;{<\;\prod_{j=1}^n {(1+\s^z_j)\over 2}\;\;e^{{H\over kT}}
\;>\over Z}\;\sim\;
{e^{{(N-n)f\over kT}}\over Z}\;=\;e^{-{nf\over kT}}
\label{asPnT}
\ee

For zero temperature we expect  Gaussian decay.

\section{Conclusion}

 We think that our work provide a link between integrable models and
chaotic models. The same mathematical apparatus appears in the
description
of both kind of models.

Let us repeat that the main result of this paper is the calculation
of $P(3)$ and $P(4)$ (\ref{P3}-\ref{P4})
by means of the multi-integral representation (\ref{intPn}).
The fact that only the logarithm $\ln{2}$ and Riemann zeta
function with odd arguments participate in the answers for
$P(1),\ldots, P(4)$ and with rational coefficients before these
functions
allows us to suppose that this is the general
property of $P(n)$. One could compare the calculation of $P(n)$ with
the many-loop calculation of the self-energy diagrams in the
renormalizable
quantum field theory which can also be expressed in terms of $\zeta$
functions
of odd arguments \cite{KR} .

Unfortunately, so far we have not got even a conjecture for $P(n)$ but
we believe that it is not an unsolvable problem. May be already
after calculation of $P(5)$ one could guess the right formula
for a generic case $P(n)$.
It would give an answer to the question discussed in the previous
section, namely, the question about the law of decay of $P(n)$
when $n$ tends to infinity.

Also it would be interesting to generalize above results to
the XXZ spin chain. Some interesting conjectures were recently
invented
by Razumov and Stroganov \cite{RS}
for the special case of the XXZ model
with $\Delta=-1/2$.
These conjectures would be supported if
it were possible to get $P(n)$ from the general integral
representation obtained by the RIMS group \cite{JMMN}.

\section{Acknowledgements}

The authors would like to thank A.~Kirillov, B. McCoy,  A.~Razumov,
M.~Shiroishi,
 Yu.~Stroganov, M.~Takahashi, L.Takhtajan and V.~Tarasov for useful discussions.
This research  has been supported by the NSF grant PHY-9988566
and by INTAS Grant no. 01-561.


\begin{thebibliography}{**}
\bibitem{KR} D. ~Kreimer,
{\it ``Knots and Feynman Diagrams''}, Cambridge University Press,
2000.
\bibitem{string}
D.J.Gross and E. Witten, Nucl. Phys. {\bf B 277}, 1, (1986);\\
M.B.Green and J.H.Schwarz, Nucl. Phys. {\bf B 181}, 502, (1981);
Nucl. Phys., {\bf B 198}, 441 , (1982);\\
J.H.Schwarz,  Phys. Rep. {\bf 89}, 223, (1982)
\bibitem{Heis} W. Heisenberg, Zeitschrift f{\"u}r Physik, {\bf vol.
49} (9-10), 619, (1928)
\bibitem{B} H.~Bethe, Zeitschrift f{\"u}r Physik, {\bf 76}, 205 (1931)
\bibitem{H} L.~Hulth\'{e}n, Ark. Mat. Astron. Fysik {\bf A 26}, 1
(1939).
\bibitem{comp} K.M.~O'Connor and W.K.~Wootters, quant-ph/0009041;\\
M.C.~Arnesen, S.~Bose and V.~Vedral, quant-ph/0009060 ;\\
 W.K.~Wootters, quant-ph/0001114
\bibitem{TS} M.Takahashi and M.Suzuki, Prog. Theor. Phys. {\bf v48} ,
2187, (1972)
\bibitem{TSK}  M. Takahashi, M. Shiroishi and A. Kl{\"u}mper, cond-
mat/0102027
\bibitem{T}  M. Takahashi,
{\it ``Thermodynamics of one-dimensional solvable models'',}\\
Cambridge University Press, 1999.
\bibitem{KIB} V.E.~Korepin, A.G.~Izergin and N.M.~Bogoliubov,
{\it ''Quantum inverse scattering method and correlation functions"},
Cambridge Univ. Press, Cambridge, 1993.
\bibitem{JMMN} M.~Jimbo, K.~Miki, T.~Miwa and A.~Nakayashiki,
{\it Phys. Lett.} {\bf A 166}, 256, (1992); hep-th/9205055.
\bibitem{KIEU} V.E.~Korepin, A.G.~Izergin, F.~Essler, D.B.~Uglov,
{\it Phys. Lett.} {\bf A 190}, 182-184, (1994).
\bibitem{T1} M.~Takahashi, J.~Phys. {\bf C 10}, 1289, (1977);
cond-mat/9708087.
\bibitem{DI} J.~Dittrich and V.I.~Inozemtsev,
``On the second-neighbor correlator in 1D XXX quantum anti-
ferromagnetic spin chain",
cond-mat/9706263.
\bibitem{BGSS} F.~Berruto, G.~Grignani, G.W.~Semenoff and P.~Sodano,
hep-th/9901142, preprint DFUPG-190-98, UBC/GS-6-98.
\bibitem{GK} J. de Gier and V.E.~Korepin, math-ph/0101036.
\bibitem{RS} A.V.~Razumov and Yu.G.~Stroganov, cond-mat/0012141;
cond-mat/0102247.
\bibitem{TIT} E.C. Titchmarch {\it ''The Theory of the Riemann Zeta-
Function  "},
Clarendon Pr, 1987.
\bibitem{R} B.Riemann, {\"U}ber die Anzahl der Primzahlen unter einer
gegebenen Gr{\"o}sse,
{\it   Monat. der Koenigl. Preuss. Akad. der Wissen. zu Berlin aus dem
Jahre 1859 (1860),
671-680; also Gesammelte mat. Werke und wissensch. Nachlass, 2 Aufl.
1892, 145-155}

\bibitem{tr}  Tanguy Rivoal
C. R. Acad. Sci. Paris Sér. I Math. 331 (2000), no. 4, 267--270

\bibitem{za} Don Zagier {\it Values of Zeta Functions and their
Applications}
First European Congress of Mathematics, Vo.II (Paris, 1992)
Prog.Math., Birkhauser, Basel-Boston,
 page 497, 1994
\bibitem{bbbl} J.M.Borwein, D. M.Bradley, D.J.Broadhurst and P.
Lisonek
{\it Special values of multiple polylogarithms}
math.CA/9910045

\bibitem{con}  J.Bost, A. Connes
Selecta Mathematica, New SEries 1 , 411-457 (1995)

\bibitem{kna}  A. Knauf
Reviews in Mathematical Physics, vol. 11, 1027-1060 (1999)

\end{thebibliography}
\end{document}